\documentclass[aps,prb,floatfix,twocolumn,showpacs]{revtex4}
\usepackage{latexsym}
\usepackage{graphicx}

\usepackage{subfigure}    
\usepackage{epsfig}
\begin{document}
\title{Quantum critical behavior driven by Hund's rule 
coupling in quantum antiferromagnets}
\author{Efstratios Manousakis}
\affiliation{Department of Physics and MARTECH,
Florida State University,
Tallahassee, FL 32306-4350, USA\\
Department of  Physics, University of Athens,
Panepistimioupolis, Zografos, 157 84 Athens, Greece
}
\date{\today}
\begin{abstract}
When localized spins on different d orbitals prefer different types 
of antiferromagnetic ordering, the
 Hund's rule coupling creates frustration. Using spin-wave 
theory we study the case of two such orbitals on a square lattice coupled 
through Hund's rule, such that the first one couples antiferromagnetically 
(AF) more strongly to its nearest  neighbors, 
while the second couples more strongly to its next nearest neighbors. 
We find that the zero temperature phase diagram has four 
regions, one characterized by
the familiar $(\pi,\pi)$ AF order, a second by the columnar $(\pi,0)$ order, a
third by a {\it canted} order and a fourth 
region where a quantum-disordered state emerges. We comment on the possible 
relevance of these
findings for the case of Fe-pnictide based antiferromagnets.

\end{abstract}
\maketitle
The FeAs layer of the parent compound of the Fe-pnictide 
superconductors\cite{JACSFeAs}
below approximately 134 K undergoes a spin-density-wave (SDW) ordering 
with a  small magnetic moment $\sim 0.35 \mu_B$ per Fe atom\cite{delacruz}.
The d-orbitals of the Fe atom are occupied by several electrons and
in the limit where the Hund's rule coupling is large compared
to the nearest neighbor (NN) and next nearest neighbor (NNN) 
antiferromagnetic couplings, which is believed to be the case
for these materials, we may expect a much larger
moment\cite{Singh,Cao,Yildirim,AF} per Fe atom ($\sim 2.6 \mu_B$). 
This, and other considerations,  have fueled a 
belief that the magnetism in these materials may be of 
itinerant type\cite{Raghu,Chubukov,Zlatko}. 
In the present paper we explore the possibility that the origin
of this significantly reduced moment is a result of frustration
introduced by the fact that the various Fe d-orbitals 
prefer different and competing type of magnetic ordering.

\begin{figure}[htp]
\includegraphics[width=2.5 in]{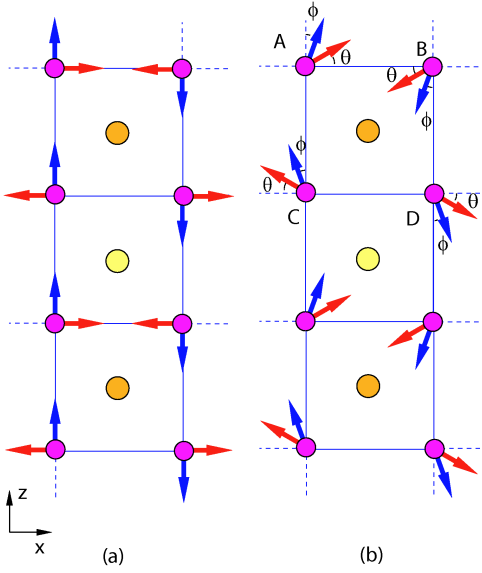}
\caption{The classical ground state of the Hamiltonian 
given by Eq.~\ref{Hamiltonian}.
(a) $J_H \to 0$.  (b) the canted state.
The red-color spins are described by ${\cal H}^{(1)}$ 
and they prefer the $(\pi,\pi)$ state,
while the blue-colored spins are described
by ${\cal H}^{(2)}$ and  prefer the columnar order.
The $J_H$ term rotates the spins towards each other by angles $\theta$ and 
$\phi$. Red circles denote Fe atoms. Orange (yellow) circles denote As atoms
above (below) the plane. }
\label{fig1}
\end{figure}

\begin{figure}[htp]
\includegraphics[width=3.0 in]{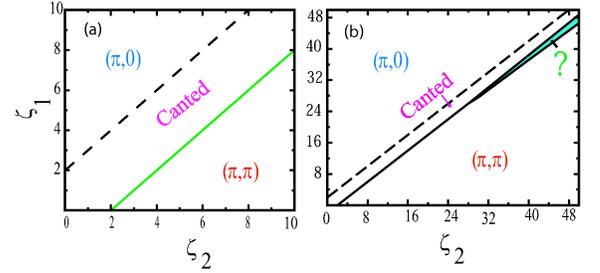}
\caption{(a): The classical phase diagram.
Here, $\zeta_{\nu} = { {S_1 S_2 J_H} \over {\alpha_{\nu}}}$
and $\alpha_1  =  2 S_1^2 (J^{(1)}_1 - 2J^{(1)}_2)$, 
$\alpha_2  =  2 S_2^2 (2J^{(2)}_2- J^{(2)}_1)$.
(b): The phase diagram as determined by the spin-wave approximation.
There is a critical value of $(\zeta^c_1,\zeta^c_2)$ 
and for $\zeta_1 > \zeta^c_1$ there is a range of $\zeta_2$ where 
the magnetic order is destroyed and a quantum-disordered phase emerges
labeled by a question-mark in the figure.}
\label{fig2}
\end{figure}

First, let us consider a simplified model in order to introduce the
reader to the problem discussed here and in order to overview our main
findings. The more realistic 
model\cite{Thimios,Thimios2}, treated within the spin-wave approximation,
will be presented below. The Hamiltonian
\begin{eqnarray}
{\cal H} &=&   J_1 \sum_{<ij>} {\bf S}_{i,1} \cdot 
{\bf S}_{j,1}
+ J_2 \sum_{<<ij>>}{\bf S}_{i,2} \cdot {\bf S}_{j,2} \nonumber \\
&-& J_H \sum_{i}
{\bf S}_{i,1} \cdot {\bf S}_{i,2},
\label{simple}
\end{eqnarray}
describes two distinct spin operators ${\bf S}_{i,1}$ and
 ${\bf S}_{i,2}$ corresponding to two different d orbitals of the same
$i^{th}$ Fe atom. The spins ${\bf S}_{i,1}$ interact antiferromagnetically 
with their NN ${\bf S}_{j,1}$ while the spins ${\bf S}_{i,2}$ 
interact with their next NN ${\bf S}_{j,2}$ (along the diagonal of the
square).  The Hund's rule coupling $J_H$ tends to align the
spins on the same atom.
The origin of qualitatively different spin-interactions
for two different d orbitals is discussed in Ref.~\onlinecite{Thimios2}.
When  $J_H=0$, the 
spins ${\bf S}_{i,1}$ order in the $(\pi,\pi)$ order indicated by the red-spins
in Fig.~\ref{fig1}(a), while the spins ${\bf S}_{i,2}$ order in the
$(\pi,0)$ (or $(0,\pi)$) order indicated by the blue-color
spins in Fig.~\ref{fig1}(a). In the absence of $J_H$ any
choice of direction of order for either type of spins is equally
acceptable, because our model is rotationally
symmetric. When $J_H>0$
the canted state of Fig.~\ref{fig1}(b) is obtained
as a compromise state between the two extremes of Fig.~\ref{fig1} (a),
by tilting the orientation of the blue spins by an angle $\phi$ toward
the red and,  the red spins toward the orientation of the
blue spins by an angle $\theta$. The two spins ``bend'' towards
each other due to Hund's rule coupling. Through this canting there is some
gain from the term proportional to $J_H$  and some loss 
due to both types of spin-spin interactions. 
When $J_1$ is not too different from 
$2 J_2$ (See Fig.~\ref{fig2}), the classical ground state
is the canted state of Fig.~\ref{fig1}(b) for any value of $J_H$. 

In this paper
we also study the role of quantum fluctuations around the classical ground 
states within spin-wave theory. We find large amplitude  
quantum spin fluctuations  when $J_1$ is sufficiently
close to $2 J_2$ and  near or in the canted phase.
Further, we find that for sufficiently large $J_H/J_1$ (and $J_H/J_2$) 
there is a quantum critical point near $J_H/J_1 \sim J_H/(2J_2) - 4$
(taking $S_1=S_2$) from where a region of a
quantum-disordered state begins.  
We discuss the consequences of our findings
for the magnetic state of the Fe-pnictides and possible
future neutron scattering experiments to search for the canted
and the disordered states.
The simpler well-known $J_1-J_2$ model\cite{Sachdev,Xu,Erica}  is
obtained from our model in the limit of very large $J_H$.
However, in order to explain the observed reduced moment,
the  $J_1-J_2$ model requires fine tuning of the $J_1/J_2$ ratio
to a value very close to the quantum critical point; on the contrary,
the present model  has a much broader parameter range yielding
large amplitude quantum spin fluctuations necessary to explain the
observed reduced moment in the Fe-pnictides. 

The problem to be discussed here is a somewhat simplified version of 
the general Hamiltonian derived in Ref.~\onlinecite{Thimios,Thimios2} 
and a generalization of the Hamiltonian given by Eq. (\ref{simple}):
\begin{eqnarray}
{\cal H} &=& \sum_{\nu} {\cal H}^{(\nu)} -J_H \sum_{i,\nu\ne\nu'}
{\bf S}_{i,\nu} \cdot {\bf S}_{i,\nu'} \\
{\cal H}^{(\nu)} &=& J^{(\nu)}_1 \sum_{<ij>} {\bf S}_{i,\nu} \cdot 
{\bf S}_{j,\nu}
+ J^{(\nu)}_2 \sum_{<<ij>>}{\bf S}_{i,\nu} \cdot {\bf S}_{j,\nu},
\label{Hamiltonian}
\end{eqnarray}
where the index $\nu=1,2,...,5$ refers to the five Fe d orbitals. 
When we consider each of the parts ${\cal H}^{(\nu)}$ separately,
if $J^{(\nu)}_2 > J^{(\nu)}_1/2$, the $(\pi,0)$ order 
(blue-colored spins of   Fig.~\ref{fig1}(a)) is stable, otherwise
within spin-wave-theory, the $(\pi,\pi)$ antiferromagnetic order 
(red-colored spins in Fig.~\ref{fig1}(a)) takes over.
For some of the Fe d orbitals in the FeAs based materials
$J^{(\nu)}_2 > J^{(\nu)}_1/2$, while for other d orbitals
this condition is not satisfied\cite{Thimios,Thimios2}. 
For simplicity, we will present here
the case of just two orbitals such that the first one, i.e., $\nu=1$,
 satisfies the condition $J^{(1)}_2 < J^{(1)}_1/2$ for $(\pi,\pi)$ order,
while the $\nu=2$ orbital satisfies the condition
$J^{(1)}_2 > J^{(1)}_1/2$ for the $(\pi,0)$ order. 

First, notice that the Hamiltonian at the classical level for certain
range of the coupling constants has
a ground state shown in Fig.~\ref{fig1} (b). 
 The blue spins of the up-sublattice
are canted by an angle $\phi$ and the red spins by an angle $\theta$ as 
shown in the figure. The total energy difference
from the energy of the state of Fig.~\ref{fig1}(a) is
$\delta E  =  - {{\alpha_1} \over 2}  \cos(2\theta) -
{{\alpha_2} \over 2}  cos(2\phi) - 
J_H S_1 S_2 sin(\theta+\phi)$,
where 
$\alpha_1  =  2 S_1^2 (J^{(1)}_1 - 2J^{(1)}_2)$, 
$\alpha_2  =  2 S_2^2 (2J^{(2)}_2- J^{(2)}_1)$,
and $S_{1,2}$ are the maximum length of the two classical spins.
In the interval  $0\le \theta \le \pi/2$, $0\le \phi \le \pi/2$, 
there are the following extrema of the energy. First,
the following two trivial solutions
$(\theta,phi) =  ({ {\pi} \over 2},0)$,
and $(\theta,\phi) = (0,{ {\pi} \over 2})$,
each of which is a stable absolute minimum, respectively, when
$\zeta_1  -  \zeta_2 > 2$, and 
$\zeta_2  -  \zeta_1  > 2,$ where 
$\zeta_{\nu} = { {S_1 S_2 J_H} \over {\alpha_{\nu}}}$.
When neither of these conditions for trivial
solutions is satisfied the stable absolute minimum is given by
\begin{eqnarray}
\sin^2(2\phi)  =  \zeta^2_2 {{1 - 
\Bigl ({{\zeta_1 - \zeta_2} \over 2 }\Bigr )^2}
\over { 1 + \zeta_1 \zeta_2}}, \hskip 0.1 in
\sin(2 \theta)  =   {{\zeta_1} \over {\zeta_2}} \sin(2 \phi).
\label{angle}
\end{eqnarray}
The classical phase diagram is shown in Fig.~\ref{fig2}(a). 
Notice that for any value of the $J_H$ there is the canted
phase with the angles given as in Eq.~\ref{angle} provided that
the other couplings $\alpha_1)$ and 
$\alpha_2$ are not very different
from each other, i.e., when they satisfy the condition discussed
above. If, however, these couplings are very different
in magnitude, the global ground state is
the one preferred by the stronger coupling, i.e., if $\alpha_2 >> \alpha_1$
the $(\pi,0)$ order is the ground state, and when $\alpha_1 >> \alpha_2$
 the $(\pi,\pi)$ state wins. 
Both transition lines separating the canted order from the 
$(\pi,\pi)$ order, or from the $(\pi,0)$ order, are lines
of second order critical points.

In order to study the role of quantum fluctuations, we 
first carry out a local rotation of the spin quantization axes
along the direction of the classical order, i.e., by angles
$\theta$ and $\phi$  for spins on sublattice A as follows:
$S^{z'}_{i,1} = \sin(\theta) S^z_{i,1} - \cos(\theta) S^x_{i,1}$, and
$S^{x'}_{i,1} = \cos(\theta) S^z_{i,1} + \sin(\theta) S^x_{i,1}$,
while the $y$ component remains unchanged, because we have assumed that
the rotation is in the $x-z$ plane (the plane of the drawing). The 
expressions
for the second component are obtained from the above by replacing
$\theta \to \pi/2 - \phi$. 
For the sublattices B, C and D, we can still use the above
expressions but with the angles $(\phi,\theta)$ replaced
by $(\pi+\phi,\pi+\theta)$, $(\pi-\phi,-\theta)$ and $(-\phi,\pi-\theta)$ 
respectively.

In order to apply the spin-wave approximation\cite{Manousakis}, we express
the operators $S^z_{i,\nu}$, $S^x_{i,\nu}$ $S^y_{i,\nu}$ using the spin deviation
operators, i.e., 
$S^z_{i,\nu} = S_{\nu} - a^{\dagger}_{i,\nu} a_{i,\nu}$,
$S^x_{i,\nu} = \sqrt{{{S_{\nu}} \over {2}}} 
(a^{\dagger}_{i,\nu} + a_{i,\nu})$
and $S^y_{i,\nu} = i \sqrt{{{S_{\nu}} \over {2}}} (a^{\dagger}_{i,\nu} - a_{i,\nu})$,
for both cases of spin ``color'' $\nu=1,2$.
By substituting these operators in the Hamiltonian given by 
Eq.~\ref{Hamiltonian}, and keeping up to quadratic terms in 
spin-deviation operators we obtain
\begin{eqnarray}
{\cal H} &=& E_{0} + \sum_{\nu,{\bf k}} \Bigl [ A^{(\nu)}_{{\bf k}} 
a^{\dagger}_{{\bf k},\nu} a_{{\bf k},\nu}  
+{{B^{(\nu)}_{{\bf k}}} \over 2 } (a^{\dagger}_{{\bf k},\nu} a^{\dagger}_{-{\bf k},\nu}
+ h.c) \Bigr ] \nonumber \\
&+& \sum_{\bf k} \Bigl ( V_{\bf k} a^{\dagger}_{{\bf k},1} a_{{\bf k},2} + h.c
+ W_{\bf k}  a^{\dagger}_{{\bf k},1} a^{\dagger}_{-{\bf k},2}
+ h.c \Bigr ),
\label{quadratic}
\end{eqnarray}
where 
\begin{eqnarray}
A^{(\nu)}_{\bf k} &=& J^{(\nu)}_1 a^{(\nu)}_1 + J^{(\nu)}_2 a^{(\nu)}_2 + 
J_H a^{(\nu)}_H, \nonumber \\
B^{(\nu)}_{\bf k} &=& J^{(\nu)}_1 b^{(\nu)}_1 + J^{(\nu)}_2 b^{(\nu)}_2,\nonumber \\
a^{(\nu)}_H &=& S_1 S_2 \sin(\phi+\theta)/S_{\nu}, \nonumber\\
V_{\bf k} &=& -{{J_H}/2} \sqrt{S_1 S_2} (1+\sin(\phi+\theta)),\nonumber\\
W_{\bf k} &=& {{J_H}/2} \sqrt{S_1 S_2}(1-\sin(\phi+\theta)),\nonumber
\label{factors}
\end{eqnarray}
where the coefficients are given in Table~\ref{table1}
and  $a_{{\bf k},\nu}$ are the Fourier components of the operators 
$a_{i,\nu}$  which are defined over the entire Brillouin Zone of 
the non-magnetically
ordered system, i.e., $-\pi < k_x,k_y \le \pi$.

\begin{table}
\caption{The factors needed in Eq.\ref{factors} are given below. The 
notation: $c_x=\cos k_x$, $c_y=\cos k_y$ and $c_{xy} = \cos k_x
\cos k_y $ is used.}
\label{table1}
\begin{tabular}{|p{30pt}|p{105pt}|p{95pt}|}\hline
 $(\nu,\mu)$ & $a^{(\nu)}_{\mu}$ &  $b^{(\nu)}_{\mu}$ \\ \hline
$(1,1)$ & $2 S_1 (2 \cos^2(\theta) + c_y \sin^2(\theta))$ &
$-2 S_1 (c_x + c_y \cos^2(\theta))$ \\ \hline
$(1,2)$ &  $4 S_1(\cos^2(\theta) c_{xy} - \cos(2 \theta))$ &
$-4 S_1 \sin^2(\theta) c_{xy}$ \\ \hline
 $(2,1)$ & $2 S_2 (2 \sin^2(\phi) + \cos^2(\phi) c_y)$ &
$-2 S_2 (c_x + c_y \sin^2(\phi))$ \\ \hline
 $(2,2)$ & $4 S_2 (\cos(2 \phi) + c_{xy} \sin^2(\phi))$ & 
$-4 S_2 \cos^2(\phi)  c_{xy} $ \\ \hline
\end{tabular}
\end{table}

There are terms proportional to 
$(S^z_{i,\nu} S^x_{j,\nu} - S^x_{i,\nu} S^z_{j,\nu})$
arising from both ${\cal H}^{(\nu)}$ and the $J_H$ term.
These terms in the spin-wave approximation lead to linear terms in
the operators $a_{i,\nu}$ and $a^{\dagger}_{i,\nu}$ and they can be 
eliminated by choosing the angles to be those
minimizing the classical energy.

Now, the quadratic Hamiltonian given by Eq.~\ref{quadratic} can be 
diagonalized by means of a canonical transformation
\begin{eqnarray}
a_{{\bf k},\nu} = \sum_{\mu=1}^2 \Bigl [ u^{(\mu)}_{{\bf k},\nu} \alpha_{{\bf k},\mu}
+ v^{(\mu)}_{{\bf k},\nu} \alpha^{\dagger}_{-{\bf k},\mu} \Bigr ],
\end{eqnarray}
where the coefficients should be chosen to preserve the
canonical commutation relations for the boson operators.
This requires the following normalization condition
$\sum_{\mu=1}^2 \Bigl [ |u^{(\mu)}_{{\bf k},\nu}|^2 -
 |v^{(\mu)}_{{\bf k},\nu}|^2 \Bigr ] = 1$.
Due to the above condition, the requirement for the canonical transformation
to transform the Hamiltonian (\ref{quadratic}) in a diagonal form as
follows
\begin{eqnarray}
{\cal H} = C + \sum_{\nu=1}^2 \omega_{{\bf k}\nu}
(\alpha^{\dagger}_{{\bf k},\nu}\alpha_{{\bf k},\nu} + { 1 \over 2} ),
\end{eqnarray}
implies that the eigenfrequencies $\omega_{{\bf k},\nu}$ 
and eigenvectors $\alpha^{\dagger}_{{\bf k},\nu}$ are given from the
set of equations $D(\omega_{{\bf k},\nu}) {\bf x}^{(\nu)} = 0$, where the matrix 
\begin{eqnarray}
D(\omega) \equiv  \left ( \begin{array}{cccc} A^{(1)}_{\bf k}-\omega
& B^{(1)}_{\bf k}  & V_{\bf k} & W_{\bf k} \\
 B^{(1)}_{\bf k} & A^{(1)}_{\bf k}+\omega
 & W_{\bf k} & V_{\bf k} \\
 V_{\bf k} & W_{\bf k} & A^{(2)}_{\bf k}-\omega & B^{(2)}_{\bf k}  \\
  W_{\bf k} & V_{\bf k} &  B^{(2)}_{\bf k} & A^{(2)}_{\bf k}+\omega
 \end{array} \right ),
\label{matrix}
\end{eqnarray}
and the components of the vector  ${\bf x}^{(\nu)}$ are 
$u^{(\nu)}_{{\bf k},1}$, $v^{(\nu)}_{{\bf k},1}$,  $u^{(\nu)}_{{\bf k},2}$, and
$v^{(\nu)}_{{\bf k},2}$. 
Here,  
we have taken advantage of the relations $(u^{(\nu)}_{-{\bf k},\mu})^*
=u^{(\nu)}_{{\bf k},\mu}$, and $(v^{(\nu)}_{-{\bf k},\mu})^*
=v^{(\nu)}_{{\bf k},\mu}$. 
We find that 
\begin{eqnarray}
\omega^2_{{\bf k},\nu} &=& \Omega_{\bf k} \pm \sqrt{\Delta_{\bf k}}, 
\end{eqnarray}
where
$\Omega_{\bf k} = {{(\eta^2_{{\bf k},1} + \eta^2_{{\bf k},2})}/2}
+ V^2_{\bf k} - W^2_{\bf k}$, $\eta^2_{{\bf k},\nu} = 
(A^{(\nu)}_{\bf k})^2 - (B^{(\nu)}_{\bf k})^2$, and
$\Delta_{\bf k} =  ( {({\eta^2_{{\bf k},1} - \eta^2_{{\bf k},2}})/2})^2  
+ (\eta^2_{{\bf k},1} + \eta^2_{{\bf k},2})(V^2_{\bf k} - W^2_{\bf k})
+ 2 (A^{(1)}_{{\bf k}}A^{(2)}_{{\bf k}} + B^{(1)}_{{\bf k}} B^{(2)}_{{\bf k}})
(V^2_{\bf k} + W^2_{\bf k}) - 4 (A^{(1)}_{{\bf k}}B^{(2)}_{{\bf k}} + 
A^{(2)}_{{\bf k}} B^{(1)}_{{\bf k}}) V_{\bf k} W_{\bf k}$.
The staggered magnetizations along the direction of the rotated
local coordinate system (by the angles $\theta$ and $\phi$)
are given by
\begin{eqnarray}
m^{\dagger}_{\nu} = S_{\nu} - {1 \over N} \sum_{\bf k} \sum_{\mu=1}^2
|v^{(\mu)}_{{\bf k},\nu}|^2.
\end{eqnarray}

\begin{figure}[htp]
\includegraphics[width=3.0 in]{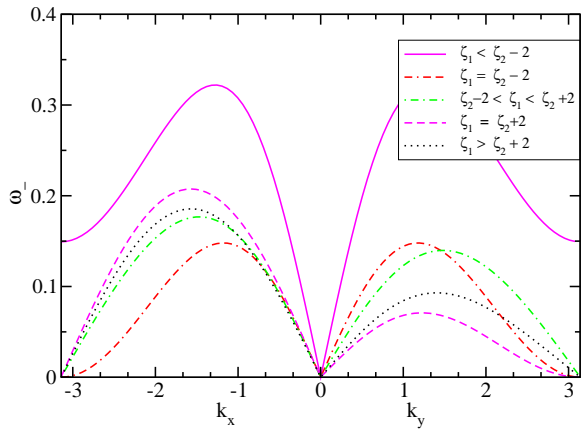}
\caption{The lowest (``acoustic'') spin-wave frequency band
along the $k_x$ (negative part of the x-axis) and $k_y$
(positive part of the x-axis) directions for various values of $\zeta_1$
which correspond to the $(\pi,\pi)$ order, the canted
phase and the $(\pi,0)$ order. These results were obtained
using $\zeta_2=4$ and $\zeta_1=1,2,3,4,6,8$.}
\label{fig3}
\end{figure}

In the following discussion and calculations presented in the figures
we restrict ourselves to the special case where $J^{(1)}_2=J^{(2)}_1=0$
and, thus, $\zeta_1 = (S_2J_H)/(2 S_1 J^{(1)}_1)$ and
$\zeta_2 = (S_1J_H)/(4 S_2 J^{(2)}_2)$.
In the entire non-magnetic BZ, there are two spin-wave frequencies,
an ``acoustic'' branch, i.e., the $\omega_{{\bf k},-}$ which 
vanishes in the long-wavelength limit and the  ``optical'' branch 
$\omega_{{\bf k},+}$ which is constant in the long-wavelength limit
and of high energy.
The acoustic frequencies are shown in Fig.~\ref{fig3} 
along the $k_x$ and $k_y$ directions keeping the value
of $\zeta_2$ constant at $\zeta_2=4$  and varying the parameter $\zeta_1$. 
For $\zeta_2=4$ there are
two critical values of $\zeta_1$, namely, $\zeta^{-}_1=2$
and $\zeta^{+}_1=6$ which define the region of the canted phase.
The spin-wave velocities along the two directions for $\zeta_1>\zeta_2-2$
are different as expected.

Notice that at the critical point $\zeta^-_1$  where, 
 we enter the canted order from
the $(\pi,\pi)$ order,  the modes at the wave vectors $(\pi,0)$ and 
$(0,\pi)$ (Figs.~\ref{fig3}) become soft. We note that 
in the pure NN antiferromagnet these modes
have maximum frequency. At the critical
point $\zeta_1=\zeta^+_1$ i.e., at border between the 
canted phase and the $(\pi,0)$ phase, these
two modes have zero frequency.

\begin{figure}[htp]
\vskip 0.3 in
\includegraphics[width=2.5 in]{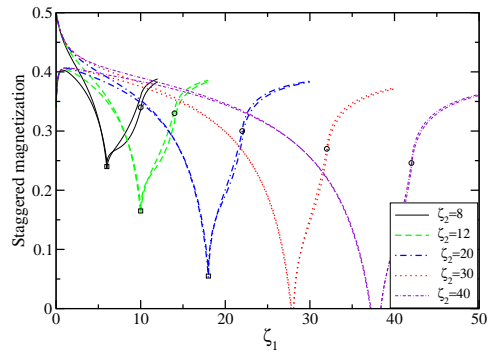}
\caption{Comparison of  staggered magnetizations $m^{\dagger}_1$ and $
m^{\dagger}_2$ for spin $S_1=S_2=1/2$ and 
for various values of $\zeta_2$ as a function of $\zeta_1$.
Notice that there is a critical value of $J_H$ where for $\zeta_1\sim 
\zeta_2-2$ none of these three forms of order survives.}
\label{fig4}
\end{figure}

In Fig.~\ref{fig4} we present the staggered magnetizations
$m^{\dagger}_1$ and $m^{\dagger}_2$ 
along the direction of order for spin $S_1=S_2=1/2$. 
The various lines are obtained by keeping $\zeta_2$ fixed and varying
$\zeta_1$.
Notice that while the magnitude of the staggered magnetization 
along the rotated direction is a continuous function across the
transition to the canted phase, there are singularities
in its derivative at $\zeta_1=\zeta^{\pm}_1 =\zeta_2\pm 2$. These singularities
indicated by the open circles and open squares 
are caused by the singularities in the abrupt change in the angles 
$\theta$ and $\phi$.

Notice that for large enough $\zeta_2$ and
for comparable value of $\zeta_1$ the staggered magnetization
along the local polarization axes becomes very small.
For large values of $\zeta_1$ the minimum occurs at $\zeta^-_1$
the boundary between the canted and the $(\pi,\pi)$ phase. 
There is a quantum critical point which is 
attained when the Hund's rule coupling is large 
compared to both $J^{(2)}_2$ and $J^{(1)}_1$.   This limit 
is believed to be the case for the Fe-pnictides. 
Our model reduces to the familiar $J_1-J_2$ model in the limit
of $J_H \to \infty$, however, as Fig.~\ref{fig4} indicates 
reaching this limit requires unrealistically large values of $J_H$ 
as compared to all other couplings.
Notice, that
the transition to the canted phase from the side of the $(\pi,0)$ order  
occurs before the staggered magnetization $m^{\dagger}_2$
becomes small, even for large values of $J_H$. 
We find that the reason for the enhancement of quantum fluctuations near the
$\zeta_1=\zeta_2-2$ boundary is that the spin-wave velocity for large $J_H$ 
decreases as we increase $J_H/J^{(1)}_1$ and this is not the case
case at the $\zeta_1=\zeta_2+2$ boundary.
Therefore, there is a quantum
disordered phase shown by the green area in Fig.~\ref{fig2}(b)
which illustrates the phase diagram as modified by quantum fluctuations.

In neutron diffraction from the FeAs based
antiferromagnets, the canted state should produce
a peak with intensity proportional to $\cos^2\phi$ at
${\bf k} = (\pi,0)$ (or $(0,\pi)$) which has 
been observed\cite{delacruz} and a peak with
low intensity proportional to $\sin^2\phi$ at ${\bf k} = (0,\pi)$
(or $(\pi,0)$).
Therefore, if the canting angle  $\phi$ is small, the latter peak might be
more difficult to resolve, and, this requires further detailed 
experimental investigation. In addition, the
magnetic unit cell of the FeAs plane of the canted phase is the same as
the structural unit cell. This is so because there is an orthorhombic
lattice distortion below 155 K and, further, the As atoms are
above and below the plane formed by the Fe atoms in a checkerboard
pattern. Therefore, diffraction using polarized neutrons
might be a simple way to probe this canted phase.
The spin-wave dispersion, which is probed by
inelastic scattering experiments\cite{Zhao,Ewings},
has no distinctly different features
from that of the $(\pi,0)$ phase (see Fig.~\ref{fig4}).

The properties of the quantum-disordered
state which emerges from the destruction of the long-range
order cannot be investigated by the present spin-wave theory approach.
As found in Refs.\onlinecite{Thimios,Yildirim,AF,Dong,Pickett}, 
the values of $J^{(\nu)}_{1}$ and $J^{(\nu)}_{2}$ are comparable, 
and, therefore, this phase may be accessible by altering these
parameters  experimentally using pressure or electron/hole doping.


\begin{thebibliography}{99}
\bibitem{JACSFeAs}Y. Kamihara, T. Watanabe, M. Hirano and H. Hosono,
J. Am. Chem. Soc.  {\bf 130}, 3296 (2008).
\bibitem{delacruz} C. de la Cruz, {\it et al.}, 
Nature {\bf 453}, 899 (2008).
\bibitem{Klauss}H.-H. Klauss, {\it et al.} 
Phys. Rev. Lett. {\bf 101} 077005
(2008).
\bibitem{Singh} D. J. Singh and M.-H. Du,  Phys. Rev. Lett. {\bf 100}, 
237003 (2008).
\bibitem{Cao}C. Cao  et al., Phys. Rev. B {\bf 77}, 220506 (2008). 
\bibitem{Yildirim} T. Yildirim, Phys. Rev. Lett. {\bf 101}, 057010(2008).
\bibitem{AF}F. Ma, Zhong-Yi Lu, Tao Xiang, arXiv:0804.3370v2.
\bibitem{Raghu} S.  Raghu et al., Phys. Rev. B {\bf 77}, 220503(2008). 
\bibitem{Chubukov} A. V. Chubukov, D. V. Efrenov and I. Eremin,  
Phys. Rev. B {\bf 78}, 134512 (2008). 
\bibitem{Zlatko} V. Cvetkovic and Z. Tesanovic,
Europhys. Lett. {\bf 85}, 37002 (2009).
\bibitem{Thimios} E. Manousakis, J. Ren, S. Meng, and E. Kaxiras, Phys. Rev. B
{\bf 78}, 205112 (2008) 
\bibitem{Thimios2} E. Manousakis, J. Ren, S. Meng, and E. Kaxiras, 
arXiv: 0902.3450.
\bibitem{Sachdev} S. Sachdev, {\it Quantum Phase Transitions}, Cambridge 
University Press, (Cambridge, U.K., 1999). 
\bibitem{Xu}C. Xu, M. Mueller, and S. Sachdev, 
Phys. Rev. B {\bf 78}, 020501(R)(2008).
\bibitem{Erica} D. -X. Yao and E. W. Carslon, Phys. Rev. B {\bf 78},
052507 (2008).
\bibitem{Manousakis} E. Manousakis, Rev. Mod. Phys. {\bf 63}, 1 (1991).
\bibitem{Zhao} J. Zhao, D.-X. Yao, S. Li, T. Hong, Y. Chen, S. Chang,
W. R. II, J. W. Lynn, H. A. Mook, G. F. Chen,
Phys. Rev. Lett. {\bf 101}, 167203 (2008).
\bibitem{Ewings} R. A. Ewings, T. G. Perring, R. I. Bewley, T. Guidi,
M. J. Pitcher, D. R. Parker, S. J. Clarke, and A. T. Boothroyd,
arXiv: 0808.2836.
\bibitem{Dong} J. Dong, H. J. Zhang, G. Xu, Z. Li, G. Li, W. Z. Hu, D. Wu, G. F. Chen, X. Dai, J. L. Luo, Z. Fang, N. L. Wang, 
Europhys. Lett., {\bf 83}, 27006 (2008).
\bibitem{Pickett}A. P. Yin, S. Leb\'egue, M. J. Han, B. P. Neal,
S. Y. Savrasov, and W. E. Pickett, Phys. Rev. Lett. {\bf 101}, 047001 (2008).
\end{thebibliography}
\end{document}